\documentclass[
a4paper,%
12pt%
]{article}

\usepackage[dvips]{graphicx}

\title{Invisible nuclear system}

\author{Sergei~P.~Maydanyuk
\thanks{E-mail: maidan@kinr.kiev.ua} \\
\small\emph{Institute for Nuclear Research,
National Academy of Sciences of Ukraine} \\
\small\emph{prosp. Nauki, 47, Kiev-28, 03680, Ukraine}}

\date{\today}

\begin{document}
\begin{sloppypar}

\maketitle

\begin{abstract}
A consecutive formalism and analysis of \emph{exactly solvable radial
reflectionless potentials with barriers}, which in the spatial
semiaxis of radial coordinate $r$ have one hole and one barrier, after
which they fall down monotonously to zero with increasing of $r$, is
presented. It has shown, that at their shape such potentials look
qualitatively like radial scattering potentials in two-partial
description of collision between particles and nuclei or radial decay
potentials in the two-partial description of decay of compound
spherical nuclear systems.
An analysis shows, that the particle propagates without the smallest
reflection and without change of an angle of motion (or tunneling)
during its scattering inside the spherically symmetric field of the
nucleus with such radial potential of interaction, i.~e. the nuclear
system with such interacting potential shows itself as
\emph{invisible} for the incident particle with any kinetic energy.
An approach for construction of a hierarchy such reflectionless
potentials is proposed, wave functions of the first potentials of
this hierarchy are found.
\end{abstract}

{\bf PACS numbers:}
11.30.Pb        
03.65.-w,       
12.60.Jv        
03.65.Xp,       
03.65.Fd,       

{\bf Keywords:}
invisible nucleus,
supersymmetry,
exactly solvable model,
reflectionless radial potentials,
inverse power potentials,
potentials of Gamov's type,
SUSY-hierarchy


\section{Introduction
\label{sec.1}}

An interest to methods of supersymmetric quantum mechanics (SUSY QM)
has been increasing every year. Initially constructed for a
description of a symmetry between bosons and fermions in field
theories, these methods during their development have formed
completely independent section in quantum mechanics
\cite{Cooper.1995.PRPLC}.

Today, the methods of SUSY QM are a powerful tool for calculation and
analysis of spectral characteristics of quantum systems, they have
shown as extremely effective in obtaining of new types of
\emph{exactly solvable potentials} and in analysis of their
properties, in an evident explanation of such unusual phenomena from
the point of view of common sense as
a \emph{resonant tunneling},
a \emph{reflectionless penetration} (or an \emph{absolute
transparency}) of the potentials (differed from the resonant tunneling
by that it exists in a whole energy spectrum, where a coefficient of
reflection is not only minimal but equals to zero also),
\emph{reinforcement of the barrier permeability} and \emph{breaking
of tunneling symmetry in opposite directions during the propagation of
multiple of particles},
\emph{absolute reflection for above-barrier energies}, 
\emph{bound states in continuous energy spectra} of systems
\cite{Zakhariev.1993.PHLTA,Zakhariev.1994.PEPAN}.

A number of papers has been increasing every year.
Here, I should like to note a fine review \cite{Cooper.1995.PRPLC}, 
to note intensively developed methods of \emph{Nonlinear (also
Polynomial, $N$-fold) supersymmetric quantum mechanics}
in~\cite{Andrianov.hep-th/9404061,%
Aoyama.quant-ph/0106037,Sato.hep-th/0109179,Gonzalez-Lopez.hep-th/0307094,%
Andrianov.2003.NuclPhys,Andrianov.2004.JPAGB,Nikitin.2002.Proc_IM}),
methods of \emph{shape invariant potentials} with different types of
parameters transformations (for example,
see~\cite{Gendenshtein.1983.JETPL,Dutt.1986.PHLTA,%
Cooper.1987.PHRVA,Dutt.1988.AJPIA,Khare.1988.JPAGB,Khare.1993.JPAGB,%
Barclay.1993.PHRVA,Balantekin.1997.PHRVA,Andrianov.quant-ph/9902057}),
methods of a description of \emph{self-similar potentials} studied by
\emph{Shabat} \cite{Shabat.1992.INPEE} and \emph{Spiridonov}
\cite{Spiridonov.1992.PRLTA,Spiridonov.hep-th/0302046} and concerned
with $q$-supersymmetry, 
methods of other types of potentials deformations and symmetries (for
example, see~\cite{Gomez-Ullate.quant-ph/0308062}),
non-stationary approaches for a description of properties and behavior
of quantum systems \cite{Samsonov.2002.Proc_IM}.
One can note papers unified methods of supersymmetry with methods of
inverse problem of quantum mechanics, and I should like to mention to
nice monography \cite{Chadan.1977} and reviews
\cite{Zakhariev.1994.PEPAN,Zakhariev.1999.PEPAN} (with a literature
list there).
An essential progress has achieved in development of the methods of
SUSY QM in spaces with different geometries
\cite{Samsonov.1997.RusPhysJ}, in non-commutative spaces
\cite{Ghosh.2005.EPJC}.
Having a powerful and universal apparatus, now the methods of SUSY QM
find their application in a number of tasks of field theories, in QCD,
in development of different models of quantum gravity, cosmology and
other.

However, in this paper I propose to pay attention into the
reflectionless phenomenon in some types of spherical symmetric quantum
systems (one note \cite{Andrianov.hep-th/9404061,Khare.1988.JPAGB,%
Chadan.1977,Nikitin.2002.Proc_IM,Bagrov.quant-ph/9804032} in
development of SUSY QM formalism for different scattering problems).
We find out a new type of radial exactly solvable reflectionless
potential, which in its shape has one hole and one barrier, after
which it falls down monotonously to zero with increasing of radial
coordinate $r$ \cite{Maydanyuk.2005.APNYA}.
Qualitatively, such potential looks like scattering potentials in
two-partial description of collision between particle and spherically
symmetric nucleus or decay potentials in the two-partial description
of decay of compound spherical nuclear system.
An analysis has shown that the particle propagates without the
smallest reflection and without change of an angle of motion (or
tunneling) in its scattering in the spherically symmetric field of the
nucleus with such radial potential of interaction, i.~e. the nuclear
system with such potential shows itself as \emph{invisible} for the
incident particle with any kinetic energy.
And this paper is devoted to an analysis of such radial reflection
potentials.

\section{SUSY-interdependence between spectral characteristics of
potentials partners in the radial problem 
\label{sec.2}}

\subsection{Darboux transformations 
\label{sec.2.1}}

Let's consider a formalism of Darboux transformations in a problem
about motion of a particle with mass $m$ in the spherically symmetric
potential field (also
see~\cite{Andrianov.hep-th/9404061,Bagrov.quant-ph/9804032}). The
spherical symmetry of the potential allows to reduce this problem to
the one-dimensional problem about the motion of this particle in the
radial field $V(r)$, defined on the positive semiaxis of $r$, where
wave function of such system looks like:
\begin{equation}
  \psi(r, \theta, \varphi) =
    \displaystyle\frac{\chi_{nl}(r)}{r}
    Y_{lm} (\theta, \varphi),
\label{eq.2.1.1}
\end{equation}
and the radial Schr\"{o}dinger equation has a form:
\begin{equation}
  H \chi_{nl}(r) =
  -\displaystyle\frac{\hbar^{2}}{2m}
  \displaystyle\frac{d^{2} \chi_{nl}(r)}{dx^{2}} +
  \biggl(V_{n}(r) +
    \displaystyle\frac{l(l+1) \hbar^{2}}{2mr^{2}} \biggr)
    \chi_{nl}(r) =
  E_{n} \chi_{nl}(r)
\label{eq.2.1.2}
\end{equation}
and differs from the one-dimensional Schr\"{o}dinger equation by a
presence of a centrifugal term. One can reduce this equation to
one-dimensional one by replacement:
\begin{equation}
  \bar{V}_{n}(r) =
    V_{n}(r) + \displaystyle\frac{l(l+1) \hbar^{2}}{2mr^{2}}.
\label{eq.2.1.3}
\end{equation}

As in the one-dimensional case, one can introduce operators $A_{1}$
and $A_{1}^{+}$:
\begin{equation}
\begin{array}{ll}

  A_{1} =
  \displaystyle\frac{\hbar}{\sqrt{2m}}
  \displaystyle\frac{d}{dr}
  + W_{1}(r), &
  A_{1}^{+} =
  -\displaystyle\frac{\hbar}{\sqrt{2m}}
  \displaystyle\frac{d}{dr}
  + W_{1}(r),
\end{array}
\label{eq.2.1.4}
\end{equation}
where $W_{1}(r)$ is a function, defined in the positive semiaxis
$0 \le r < +\infty$ and continuous in it with an exception of some
possible points of discontinuity. Then one can determine an
interdependence between two hamiltonians of the propagation of the
particle with mass $m$ in the fields $\bar{V}_{1}(r)$ and
$\bar{V}_{2}(r)$:
\begin{equation}
\begin{array}{l}
  H_{1} = A_{1}^{+} A_{1} + C_{1} =
  -\displaystyle\frac{\hbar^{2}}{2m}
  \displaystyle\frac{d^{2}}{dr^{2}}
  + \bar{V}_{1}(r), \\
  H_{2} = A_{1} A_{1}^{+} + C_{1} =
  -\displaystyle\frac{\hbar^{2}}{2m}
  \displaystyle\frac{d^{2}}{dr^{2}}
  + \bar{V}_{2}(r),
\end{array}
\label{eq.2.1.5}
\end{equation}
where each potential is expressed through one function $W_{1}(r)$:
\begin{equation}
\begin{array}{ll}
  \bar{V}_{1}(r) =
  W_{1}^{2}(r) - \displaystyle\frac{\hbar}{\sqrt{2m}}
  \displaystyle\frac{d W_{1}(r)}{dr} + C_{1}, &
  \bar{V}_{2}(r) =
  W_{1}^{2}(x) + \displaystyle\frac{\hbar}{\sqrt{2m}}
  \displaystyle\frac{d W_{1}(r)}{dr} + C_{1}.
\end{array}
\label{eq.2.1.6}
\end{equation}
One can find:
\begin{equation}
  \bar{V}_{2} (r) - \bar{V}_{1}(r) =
  V_{2} (r) - V_{1}(r) =
    2\displaystyle\frac{\hbar}{\sqrt{2m}}
    \displaystyle\frac{d W_{1}(r)}{dr}.
\label{eq.2.1.7}
\end{equation}
The determination of the potentials $V_{1}(r)$ and $V_{2}(r)$ of two
quantum systems on the basis of one function $W_{1}(r)$ establishes
the interdependence between spectral characteristics (spectra of
energy, wave functions, S-matrixes) of these systems. We shall
consider this interdependence, as the interdependence given by
Darboux transformations in the radial problem,
and we shall name $W_{1}(r)$ as \emph{superpotential}, potentials
$V_{1}(r)$ and $V_{2}(r)$ as \emph{supersymmetric potentials-partners}.

Note, that there is a constant $C_{1}$ in the definition
(\ref{eq.2.1.2}) of the hamiltonians of two quantum systems.
If to choose $C_{1}=E^{(1)}_{0}, E^{(2)}_{0} \ne E^{(1)}_{0}$
($E^{(1)}_{0}$ and $E^{(2)}_{0}$ are the lowest levels of energy
spectra of the first and second hamiltonians $H_{1}$ and $H_{2}$),
then we obtain the most widely used construction two hamiltonians
$H_{1}$ and $H_{2}$ in the one-dimensional case on the basis of the
operators $A_{1}$ and $A_{1}^{+}$ (for example, see p.~287--289
in~\cite{Cooper.1995.PRPLC}).
However, this case corresponds to bound states in the discrete regions
of the energy spectra of two studied quantum systems. For study of
scattering, decay or synthesis processes in the radial consideration
usually we deal with unbound states with the continuous region of the
energy spectra (with the lowest energy levels
$C_{1} = E^{(1)}_{0} = E^{(2)}_{0} = 0$) of quantum systems.
Therefore, one need to use $C_{1}=0$ for obtaining the interdependence
between the spectral characteristics of two systems on the basis of
Darboux transformations (and we obtain a construction of hierarchy of
potentials as in~\cite{Maydanyuk.2005.APNYA}, see p.~443--445):
\begin{equation}
\begin{array}{l}
  H_{1} = A_{1}^{+} A_{1} =
  -\displaystyle\frac{\hbar^{2}}{2m}
  \displaystyle\frac{d^{2}}{dr^{2}}
  + \bar{V}_{1}(r), \\
  H_{2} = A_{1} A_{1}^{+} =
  -\displaystyle\frac{\hbar^{2}}{2m}
  \displaystyle\frac{d^{2}}{dr^{2}}
  + \bar{V}_{2}(r).
\end{array}
\label{eq.2.1.8}
\end{equation}

\subsection{The interdependence between wave functions 
\label{sec.2.3}}

We shall study two quantum systems, in each of which there is the
scattering of the particle on the potential $V_{1}(r)$ or $V_{1}(r)$.
Further, we shall not consider processes, concerned with loss of
complete energy of systems (for example, dissipation, bremsstrahlung
etc.). The energy spectra of these systems are \emph{continuous}, and
their lowest levels are \emph{zero}.
In accordence with (\ref{eq.2.1.8}), we write:
\begin{equation}
\begin{array}{l}
  H_{1} \chi^{(1)}_{k,l} =
  A_{1}^{+} A_{1} \chi^{(1)}_{k,l} =
  E^{(1)}_{k,l} \chi^{(1)}_{k,l}, \\
  H_{2} \chi^{(2)}_{k^{\prime},l^{\prime}} =
  A_{1} A_{1}^{+} \chi^{(2)}_{k^{\prime},l^{\prime}} =
  E^{(2)}_{k^{\prime},l^{\prime}} \chi^{(2)}_{k^{\prime},l^{\prime}},
\end{array}
\label{eq.2.3.1}
\end{equation}
where $E^{(1)}_{k, l}$ and $E^{(2)}_{k^{\prime}, l^{\prime}}$ are the
energy levels of two systems with orbital quantum numbers $l$ and
$l^{\prime}$,
$\chi^{(1)}_{k,l}(x)$ and $\chi^{(2)}_{k^{\prime}, l^{\prime}}(x)$ are
the radial components of wave functions, concerned with these levels,
$k = \displaystyle\frac{1}{\hbar}\sqrt{2mE^{(1)}_{k,l}}$ and
$k^{\prime} =
\displaystyle\frac{1}{\hbar}\sqrt{2mE^{(2)}_{k^{\prime},l^{\prime}}}$
are wave vectors corresponding to the levels $E^{(1)}_{k,l}$ and
$E^{(2)}_{k^{\prime}, l^{\prime}}$.
From (\ref{eq.2.3.1}) we obtain:
\begin{equation}
  H_{2} (A_{1} \chi^{(1)}_{k,l}) =
  A_{1} A_{1}^{+} (A_{1} \chi^{(1)}_{k,l}) =
  A_{1} (A_{1}^{+} A_{1} \chi^{(1)}_{k,l}) =
  A_{1} (E^{(1)}_{k,l} \chi^{(1)}_{k,l}) =
  E^{(1)}_{k,l} (A_{1} \chi^{(1)}_{k,l}).
\label{eq.2.3.2}
\end{equation}
We see, that the function $f(r)=A_{1} \chi^{(1)}_{k,l}(r)$ is the
eigen-function of the operator $\hat{H}_{2}$ with quantum number $l$
to a constant factor, i.~e. it represents the wave function
$\chi^{(2)}_{k^{\prime},l}(r)$ of the hamiltonian $H_{2}$. The energy
level $E^{(1)}_{k,l}$ must be the eigen-value of this operator
exactly, i.~e. it represents the energy level $E^{(2)}_{k^{\prime},l}$
of this hamiltonian. Here, new wave function and energy level have the
same index ${k^{\prime}}$. One can write:
\begin{equation}
\begin{array}{lcr}
  A_{1} \chi^{(1)}_{k,l} (r) =
    N_{2} \chi^{(2)}_{k^{\prime},l} (r), &
  E^{(1)}_{k,l} = E^{(2)}_{k^{\prime},l}, &
  N_{2} = const.
\end{array}
\label{eq.2.3.3}
\end{equation}

Taking into account (\ref{eq.2.3.1}), one can write:
\begin{equation}
  H_{1} (A_{1}^{+} \chi^{(2)}_{k^{\prime},l^{\prime}}) =
  A_{1}^{+} A_{1} (A_{1}^{+} \chi^{(2)}_{k^{\prime},l^{\prime}}) =
  A_{1}^{+} (A_{1} A_{1}^{+} \chi^{(2)}_{k^{\prime},l^{\prime}}) =
  A_{1}^{+} (E^{(2)}_{k^{\prime},l^{\prime}}
    \chi^{(2)}_{k^{\prime},l^{\prime}}) =
  E^{(2)}_{k^{\prime},l^{\prime}}
    (A_{1}^{+} \chi^{(2)}_{k^{\prime},l^{\prime}}).
\label{eq.2.3.4}
\end{equation}
and obtain:
\begin{equation}
\begin{array}{lcr}
  A_{1}^{+} \chi^{(2)}_{k^{\prime},l^{\prime}} (r) =
    N_{1} \chi^{(1)}_{k,l^{\prime}} (r), &
  E^{(2)}_{k^{\prime},l^{\prime}} = E^{(1)}_{k,l^{\prime}}, &
  N_{1} = const.
\end{array}
\label{eq.2.3.5}
\end{equation}

Thus, we obtain the following interdependences between the wave
functions and the levels of the continuous energy spectra of two
systems SUSY-partners in the radial problem:
\begin{equation}
\begin{array}{cccc}
  \chi^{(1)}_{k,l^{\prime}} (r) =
    \displaystyle\frac{1}{N_{1}}
    A_{1}^{+} \chi^{(2)}_{k^{\prime},l^{\prime}} (r), &
  \chi^{(2)}_{k^{\prime},l} (r) =
    \displaystyle\frac{1}{N_{2}}
    A_{1} \chi^{(1)}_{k,l} (r), &
  E^{(1)}_{k,l} = E^{(2)}_{k^{\prime},l}, &
  E^{(1)}_{k,l^{\prime}} = E^{(2)}_{k^{\prime},l^{\prime}}.
\end{array}
\label{eq.2.3.6}
\end{equation}
Darboux transformations establish the interdependence between the
wave functions for the same energy levels of two systems. The
coefficients $N_{1}$ and $N_{2}$ can be calculated from a
normalization conditions for the wave functions (for the continuous
energy spectra), and boundary condition are defined by scattering or
decay process.

\subsection{The interdependence between amplitudes of transittion
and reflection
\label{sec.2.4}}

For scattering the radial superpotential $W_{1}(r)$, the potentials
$V_{1}(r)$ and $V_{2}(r)$ are finite in the whole spatial region of
their definition and in asymptotic they tend to zero:
\begin{equation}
\begin{array}{ll}
  W_{1} (r \to +\infty) = 0, &
  V_{1} (r \to +\infty) = V_{2} (r \to +\infty) = 0.
\end{array}
\label{eq.2.4.1}
\end{equation}
Let's find an interdependence between resonant and potential
components of S-matrixes of these systems (for example,
also see~\cite{Andrianov.hep-th/9404061}).

One can describe the particle motion in the direction to zero inside
the fields $V_{1}(r)$ and $V_{2}(r)$ with use of plane waves
$e^{-ikr}$ (we assume, that the plane waves of both systems have the
same wave vectors $k$).
In spatial asymptotic regions we obtain transmitted waves
$T_{1}(k)e^{ikx}$ and $T_{2}(k)e^{ikx}$, which are formed in result of
total propagation (with possible tunneling) through the potentials and
describe the resonant scattering of the particle on the potentials,
and reflected waves $R_{1}(k)e^{ikx}$ and $R_{2}(k)e^{ikx}$, which are
formed in result of reflection from the potentials and describe the
potential scattering of the particle on the potentials.
For each process of scattering one can write components of wave
functions, which are formed in result of the transmission through the
potential and the reflection from it:
\begin{equation}
\begin{array}{ll}
  \chi_{inc+ref}^{(1)}(k, r \to +\infty) =
    \bar{N}_{1} (e^{-ikr} + R_{1} e^{ikr}), &
  \chi_{tr}^{(1)}(k, r \to +\infty) \to
    \bar{N}_{1} T_{1} e^{ikr}, \\
  \chi_{inc+ref}^{(2)}(k, r \to +\infty) =
    \bar{N}_{2} (e^{-ikr} + R_{2} e^{ikr}), &
  \chi_{tr}^{(2)}(k, r \to +\infty) \to
    \bar{N}_{2} T_{2} e^{ikr},
\end{array}
\label{eq.2.4.2}
\end{equation}
where the coefficients $\bar{N}_{1}$ and $\bar{N}_{2}$ can be found
from the normalization conditions.

Using (\ref{eq.2.4.2}) for the wave functions in asymptotic region,
taking into account the interdependence (\ref{eq.2.3.6}) between them
and definitions (\ref{eq.2.1.1}) for the operators $A_{1}$ and
$A_{1}^{+}$, we obtain:
\begin{equation}
\begin{array}{l}
  \bar{N}_{1} \biggl( e^{-ikr} + R_{1} e^{ikr} \biggr) =
    \displaystyle\frac{\bar{N}_{2}}{N_{1}}
    \displaystyle\frac{ik\hbar}{\sqrt{2m}}
    \biggl(e^{-ikr} - R_{2} e^{ikr} \biggr), \\
  \bar{N}_{1} T_{1} e^{ikr} =
    -\displaystyle\frac{\bar{N}_{2}}{N_{1}} 
    \displaystyle\frac{ik\hbar}{\sqrt{2m}} T_{2} e^{ikr}.
\end{array}
\label{eq.2.4.3}
\end{equation}
These expressions are carried out only, if items with the same
exponents are equal between themselves. We find:
\begin{equation}
  \bar{N}_{1} =
    \displaystyle\frac{\bar{N}_{2}}{N_{1}}
    \displaystyle\frac{ik\hbar}{\sqrt{2m}}
\label{eq.2.4.4}
\end{equation}
and 
\begin{equation}
\begin{array}{cc}
  R_{1}(k) = - R_{2}(k), &
  T_{1}(k) = - T_{2}(k).
\end{array}
\label{eq.2.4.5}
\end{equation}

Exp.~(\ref{eq.2.4.5}) establish the interdependence between the
amplitudes of the transmission $T_{1}(k)$, $T_{2}(k)$ and the
amplitudes of the reflection $R_{1}(k)$, $R_{2}(k)$ for the particle
relatively two potentials. Squares of modules of the transmitted and
reflected amplitudes represent the resonant and potential components
of the S-matrixes for two systems.
We see, that all these values do not depend on the normalized
coefficients $N_{1}$, $N_{2}$, $\bar{N}_{1}$, $\bar{N}_{2}$.

One can introduce the matrix of scattering $S_{l}(k)$ for $l$-partial
wave:
\begin{equation}
  \chi_{nl}(r) \sim S_{l}(k) e^{ikr} - (-1)^{l} e^{-ikr}
\label{eq.2.4.6}
\end{equation}
and determine a phase shift $\delta_{l}(k)$:
\begin{equation}
  e^{i\delta_{l}(k)} = S_{l}(k).
\label{eq.2.4.7}
\end{equation}
Then with taking into account (\ref{eq.2.4.2}), we find:
\begin{equation}
  S_{l}(k) = (-1)^{l+1} (R_{l}(k) + T_{l}(k)).
\label{eq.2.4.8}
\end{equation}
One can see, how these partial components of the S-matrixes and the
phases for two systems are interdependent (also
see~\cite{Cooper.1995.PRPLC}, p.~278--279):
\begin{equation}
\begin{array}{cc}
  S_{l}^{(1)}(k) = - S_{l}^{(2)}(k), &
  \delta_{l}^{(1)}(k) = \delta_{l}^{(2)}(k) + \pi/2.
\end{array}
\label{eq.2.4.9}
\end{equation}

Let's consider a spherically symmetric quantum system with the radial
potential, to which a zero amplitude of the reflection $R(k)$ of the
wave function corresponds. The particle during its scattering
in this field propagates into a center without the smallest
reflection by the field. In particular, such is a nul radial
potential. We shall name such quantum systems and their radial
potentials as \emph{reflectionless} or \emph {absolutely
transparent}.
Then from (\ref{eq.2.4.5}) one can see, that the potential-partner for
the reflectionless potential is reflectionless also in that region,
where it is finite. If such potential is finite on the whole region
of its definition, then it is reflectionless completely (i.~e. in
standard definition of quantum mechanics). A series of the finite
potentials of hierarchy, which contains the nul radial potential,
should be reflectionless also. Using this simple idea and knowing a
form of only one reflectionless potential, one can construct many new
exactly solvable radial reflectionless potentials.

\section{Spherically symmetric systems with absolute transparency
\label{sec.3}}

\subsection{A radial reflectionless potentials with barriers 
\label{sec.3.1}}

In~\cite{Maydanyuk.2005.APNYA} (see sec.~5.3.2, p.~459--462) an
one-dimensional superpotential, defining a reflectionless potential
which in semiaxis $0 < x < +\infty$ has one hole, one barrier and
then with increasing of $x$ falls down monotonously to zero in
asymptotic region, had found.
As this superpotential is obtained on the basis of interdependence
between two one-dimensional hamiltonians with continuous energy
spectra, one can use it in the problem about scattering of a particle
in the spherically symmetric field with a barrier and with orbital
quantum number $l=0$. In such case, we have:
\begin{equation}
\begin{array}{ll}
  W(r) =
      \displaystyle\frac{2\beta - \alpha}{f(\bar{r})} -
        \displaystyle\frac{\beta}{\bar{r}}, &
        \mbox{при } 2\beta \ne \alpha,
\end{array}
\label{eq.3.1.1}
\end{equation}
where
\begin{equation}
  f(\bar{r}) = C(2\beta - \alpha) \bar{r}^{2\beta / \alpha} +
               \bar{r}.
\label{eq.3.1.2}
\end{equation}
Here $\bar{r} = r+r_{0}$,
$\beta$ and $C$ are arbitrary real positive constants,
$r_{0}$ is a positive number close to zero, and a designation 
$\alpha = \displaystyle\frac{\hbar}{\sqrt{2m}}$ is introduced. 
This superpotential is defined on the positive semiaxis of $r$ 
(at $r > r_{0}$).

Let's find potentials-partners for the superpotential (\ref{eq.3.1.1}).
In accordance with (\ref{eq.2.1.6}), we obtain:
\begin{equation}
\begin{array}{lcl}
  V_{1,2}(r) & = &
    \displaystyle\frac{(2\beta - \alpha)^{2}} {f^{2}(\bar{r})} -
    \displaystyle\frac{2\beta (2\beta - \alpha)}
      {\bar{r} f(\bar{r})} +
    \displaystyle\frac{\beta^{2}}{\bar{r}^{2}} \pm \\

    & \pm &
    \Biggl(
    \displaystyle\frac{(2\beta - \alpha)^{2}} {f^{2}(\bar{r})} -
    \displaystyle\frac{2\beta (2\beta - \alpha)}
      {\bar{r} f(\bar{r})} +
    \displaystyle\frac{\alpha\beta}{\bar{r}^{2}}
    \Biggr)
\end{array}
\label{eq.3.1.3}
\end{equation}
or
\begin{equation}
\begin{array}{lcl}
  V_{1}(r) & = &
    \displaystyle\frac{\beta (\beta - \alpha)} {\bar{r}^{2}}, \\
  V_{2}(r) & = &
    2\displaystyle\frac{(2\beta - \alpha)^{2}} {f^{2}(\bar{r})} -
    \displaystyle\frac{4\beta (2\beta-\alpha)}
      {\bar{r} f(\bar{r})} +
    \displaystyle\frac
    {\beta (\beta+\alpha)} {\bar{r}^{2}}.
\end{array}
\label{eq.3.1.4}
\end{equation}

From (\ref{eq.3.1.4}) one can see that at $\beta = \alpha$ the first
potential $V_{1}(r)$ obtains zero value and, therefore, it becomes
reflectionless. Then, according to (\ref{eq.2.4.5}), if the second
potential $V_{2}(r)$ is finite in a whole region of its definition,
then it should be reflectionless also.
At $\beta = \alpha$ we obtain:
\begin{equation}
  V_{2}(r) =
    \displaystyle\frac{2\alpha^{2}}
    {\biggl(r + r_{0} + \displaystyle\frac{1}{C\alpha}\biggr)^{2}}.
\label{eq.3.1.5}
\end{equation}
We see, that this potential is finite in the whole region of its
definition at any values of the parameters $C>0$ and $r_{0} \ge 0$.
Thus, we have obtained the reflectionless potential of the inverse
power type with a shift to the left, which is defined on the whole
positive semiaxis of $r$ (including $r=0$ and
$r_{0}=0$).

In accordance with \cite{Maydanyuk.2005.APNYA} (see p.~452--455,
sec.~5.1.2), one can construct a hierarchy of the inverse power
potentials, and a general solution of the potential with arbitrary
number $n$ can be written down so:
\begin{equation}
\begin{array}{ll}
  V_{n}(r) =
    \displaystyle\frac{\gamma_{n} \alpha^{2}}{\bar{r}^{2}}, &
  \gamma_{n \pm 1} = 1 + \gamma_{n} \pm \sqrt{4\gamma_{n}+1}.
\end{array}
\label{eq.3.1.6}
\end{equation}
If to require, that the first potential $V_{1}(r)$ in this hierarchy
must be constant (i.~e. at $\gamma_{1}=0$ and $n=1$), then all
hierarchy of the inverse power potentials (\ref{eq.3.1.6}) becomes
the \emph{hierarchy of the reflectionless inverse power potentials},
and the solution (\ref{eq.3.1.5}) becomes the general solution for
the reflectionless inverse power potential.
Note, that \emph{when the hierarchy of the inverse power potentials
becomes reflectionless, then the coefficients $\gamma_{n}$ become
integer numbers}. We write its first values:
\begin{equation}
  \gamma_{n} = 0, 2, 6, 12, 20, 30, 42...
\label{eq.3.1.7}
\end{equation}

Now, if to calculate $\beta_{n}$ for given $\gamma_{n}$ with number
$n$ from (\ref{eq.3.1.7}) from the following condition:
\begin{equation}
  \beta_{n} (\beta_{n}-\alpha) = \gamma_{n} \alpha^{2},
\label{eq.3.1.8}
\end{equation}
then the first potential $V_{1}(r)$ from (\ref{eq.3.1.4}) becomes
reflectionless inverse power potential (at $\beta =\beta_{n}$). The
second potential $V_{2}(r)$ from (\ref{eq.3.1.4}) is finite in the
whole region of its definition (including $r=0$) and should be
reflectionless also, however it is not inverse power potential. So,
substituting the coefficients $\gamma_{n}$ with other numbers $n$
into the second expression (\ref{eq.3.1.4}) for the potential
$V_{2}(r)$, one can construct the whole hierarchy of the radial
reflectionless potentials of this new type.

\begin{figure}[ht]
\centerline{
\includegraphics[width=57mm]{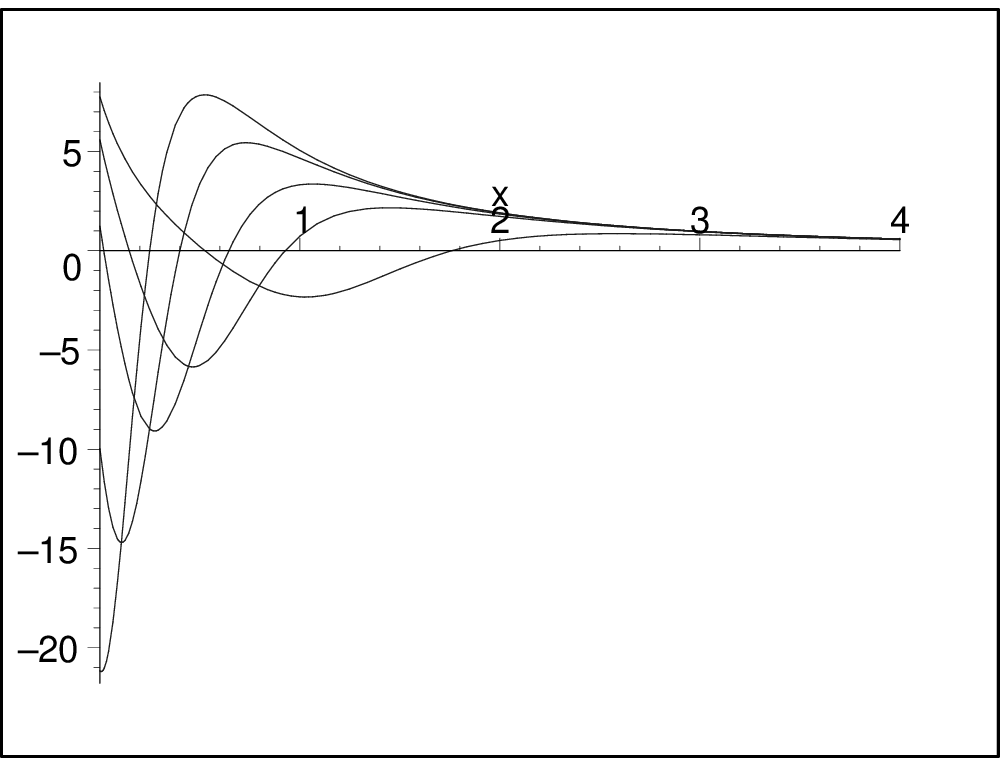}
\includegraphics[width=57mm]{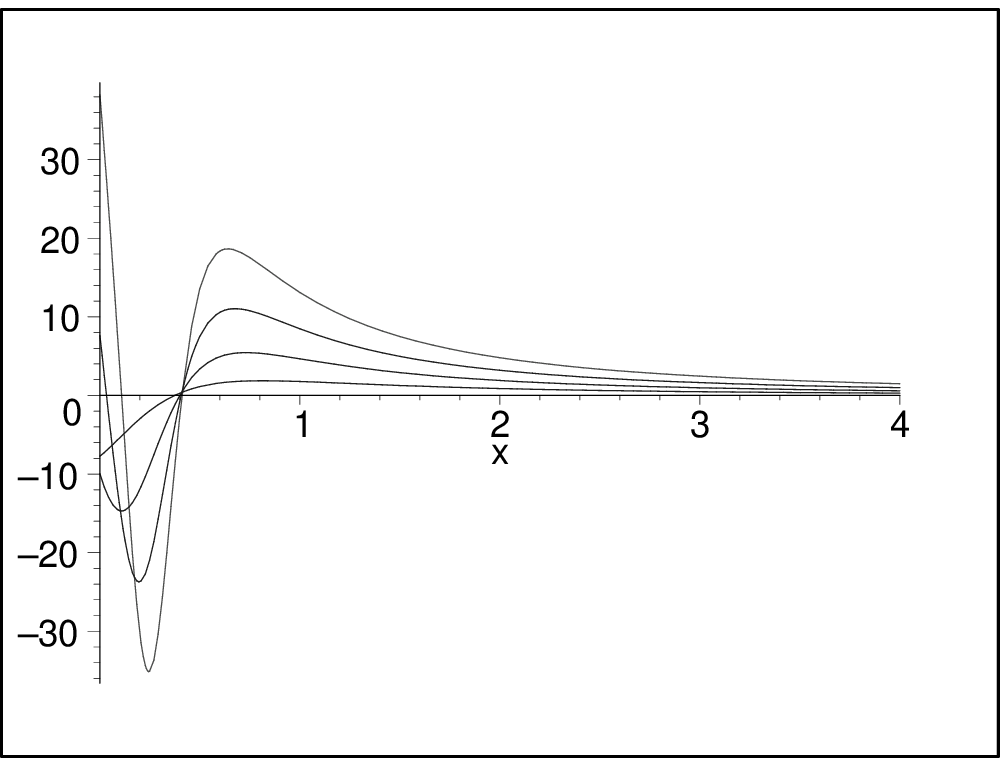}} 
\caption{
A dependence of the radial potential $V_{2}(r)$ on
$C$ and $\gamma_{n}$:
(a) the barrier maximum and the hole minimum of this potential
are changed along the axis $r$ at the change of $C$ 
(at $C = 0.01, 0.1, 0.3, 1.0, 2.5$, $\gamma_{n}=6$, $r_{0}=0.5$);
(b) the barrier maximum of this potential practically does not
changed along the axis $r$ at the change of $\gamma_{n}$ 
(at $C = 1$, $\gamma_{n}=2, 6, 12, 20 $, $r_{0}=0.5$).
\label{fig.1}}
\end{figure}

\begin{figure}[ht]
\centerline{
\includegraphics[width=80mm]{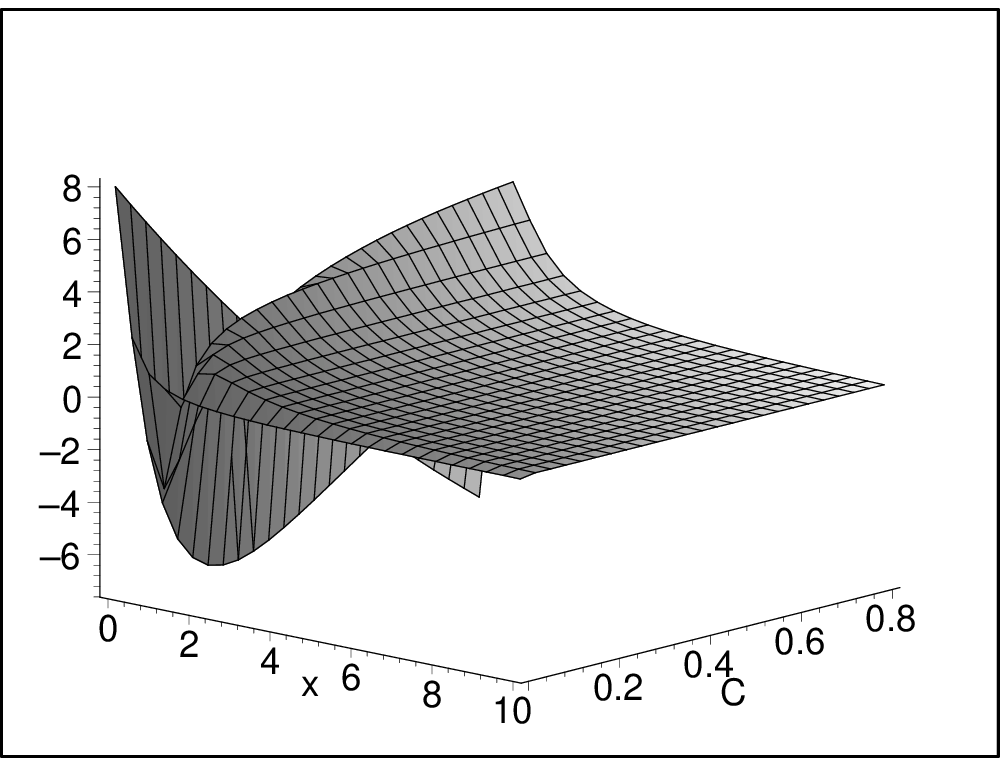}
\includegraphics[width=80mm]{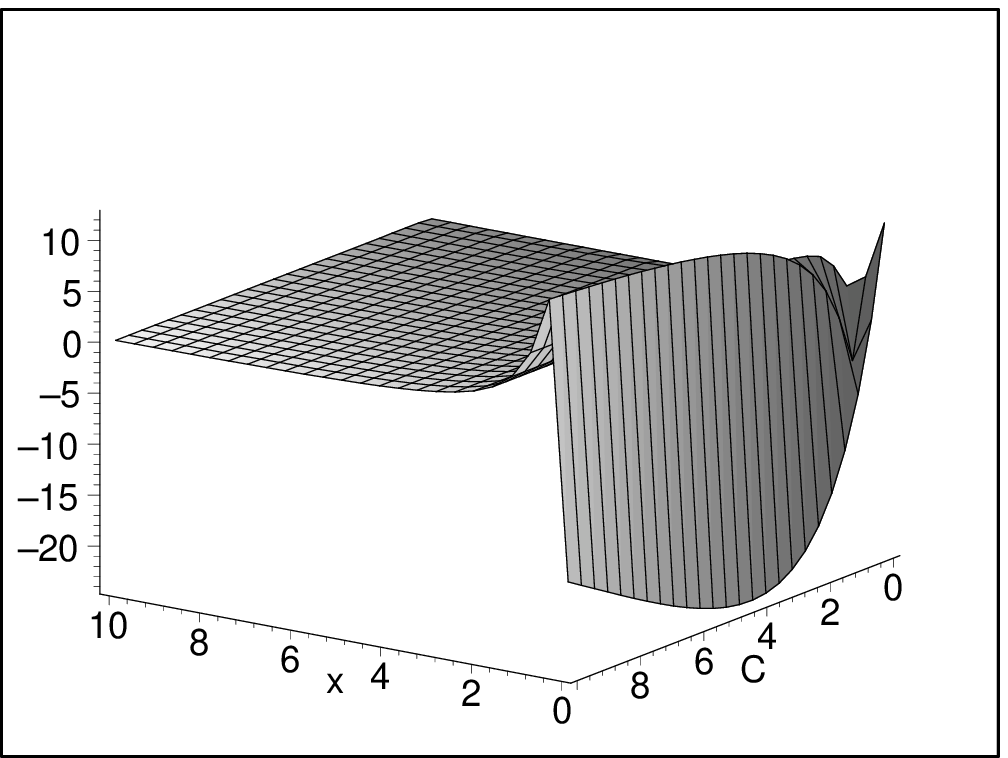}} 
\caption{The reflectionless radial exactly solvable potential
$V_{2}(r)$ with the barrier.
Continuous change of its shape at variation of $C$
($\gamma_{n}=6$, $r_{0}=0.5$)
\label{fig.2}}
\end{figure}

In Fig.~\ref{fig.1} the potential $V_{2}(r)$ for the chosen values of
the parameters $C$ and $\gamma_{n}$ is shown. From here one can see,
that such potential has one hole and one barrier, after which it
falls down monotonously to zero with increasing of the radial
coordinate $r$.
In its behavior such potential looks qualitatively like radial
potentials with barriers used in theory of nuclear collisions for
a description of scattering of particles on spherical nuclei, and
for a description of decay and synthesis of nuclei of a spherical
type also.
This potential is reflectionless, if the parameter $\gamma_{n}$ has
discrete values from the sequence (\ref{eq.3.1.7}).
For any reflectionless potential with given $\gamma_{n}$ one can
displace continuously its barrier and hole along an axis $r$ by use
of the parameter $C$. Such deformation of the shape of the
reflectionless potential is shown in Fig.~\ref{fig.2}.

\subsection{An analysis of wave functions 
\label{sec.3.2}}

\subsubsection{Wave functions for the reflectionless inverse power
potential
\label{sec.3.2.1}}

Let's find a radial wave function describing the scattering of the
particle on the inverse power reflectionless potential
(\ref{eq.3.1.5}) at $l=0$.
For the potential $V_{1}(r)$ with zero value of (\ref{eq.3.1.4}) one
can write its radial wave function (for arbitrary energy level
concerned with wave vector $k$) at $l=0$ by such a way
(at $\beta=\alpha$):
\begin{equation}
  \chi_{l=0}^{(1)}(k,r) =
    \bar{N}_{1} (e^{-ikr} - S_{l=0}^{(1)} e^{ikr}).
\label{eq.3.2.1.1}
\end{equation}
Then, one can find a radial wave function at $l=0$ for the
reflectionless potential $V_{2}(r)$ of (\ref{eq.3.1.5}) (for the
energy level corresponding to the wave vector $k$) on the basis of
the second expression of (\ref{eq.2.3.6}). Taking into account
(\ref{eq.2.1.4}) and (\ref{eq.2.4.9}), we obtain:
\begin{equation}
\begin{array}{lcl}
  \chi_{l=0}^{(2)}(k,r) & = &
      \displaystyle\frac{1}{N_{2}} A_{1} \chi_{l=0}^{(1)}(k,r) =
      \displaystyle\frac{\bar{N}_{1}}{N_{2}}
      \biggl( \alpha\displaystyle\frac{d}{dr} + W(r) \biggr)
      \Bigl(e^{-ikr} - S_{l=0}^{(1)} e^{ikr}\Bigl) = \\
    & = &
      \chi_{l=0}^{(-)}(k,r) - S_{l=0}^{(2)}\chi_{l=0}^{(+)}(k,r),
\end{array}
\label{eq.3.2.1.2}
\end{equation}
where
\begin{equation}
  \chi_{l=0}^{(\pm)}(k,r) =
      \bar{N}_{2}
        \biggl( 1 \mp \displaystyle\frac{iW(r)}{k\alpha} \biggr)
        e^{\pm ikr} 
\label{eq.3.2.1.3}
\end{equation}
and
\begin{equation}
  \bar{N}_{2} = i\displaystyle\frac{\bar{N}_{1}}{k\alpha N_{2}}
\label{eq.3.2.1.4}
\end{equation}

In accordance with main statements of quantum mechanics, for
applying such form of the radial wave function to the description
of scattering of the particle in the field of the potential
$V_{2}(r)$, it needs to achieve a boundary requirement
$\chi_{l=0}^{(2)}(k,r) \to 0$ at $r \to 0$, which gives a finiteness
of the wave function (\ref{eq.2.1.1}) at $r=0$ (and $S_{l=0}^{(2)}$
must have finite values and be not zero). One can see from
(\ref{eq.3.2.1.2}), that it is fulfilled only in case ($W(r)$ is
real):
\begin{equation}
\begin{array}{lcl}
  Re (S_{l=0}^{(2)}) =
    \displaystyle\frac{k^{2}\alpha^{2}-W^{2}(0)}
    {k^{2}\alpha^{2}+W^{2}(0)}, &
  Im (S_{l=0}^{(2)}) =
    \displaystyle\frac{2W(0)k\alpha}{k^{2}\alpha^{2}+W^{2}(0)},
\end{array}
\label{eq.3.2.1.5}
\end{equation}
where
\begin{equation}
  W(0) =
    -\displaystyle\frac{\alpha}{r_{0} + \displaystyle\frac{1}{C\alpha}}.
\label{eq.3.2.1.6}
\end{equation}
For the partial components of the S-matrix the following property
$|S_{l=0}|^{2} = 1$ is fulfilled also. In limit $r \to 0$ we obtain
the following expression for the radial wave function:
\begin{equation}
  \chi_{l=0}^{(2)}(k,r) =
    \bar{N}_{2} \biggl(1 + \displaystyle\frac{iW(0)}{k\alpha}\biggr)
    \biggl(e^{-ikr} - S_{l=0}^{(2)}
      \displaystyle\frac{k\alpha - iW(0)}{k\alpha + iW(0)}
      e^{ikr}\biggl),
\label{eq.3.2.1.7}
\end{equation}
which in its form coincides with Exp.~(\ref{eq.3.2.1.1}) for the
wave function for the potential $V_{1}(r)$ from (\ref{eq.3.1.4})
with zero value.

\begin{figure}[ht]
\centerline{
\includegraphics[width=50mm]{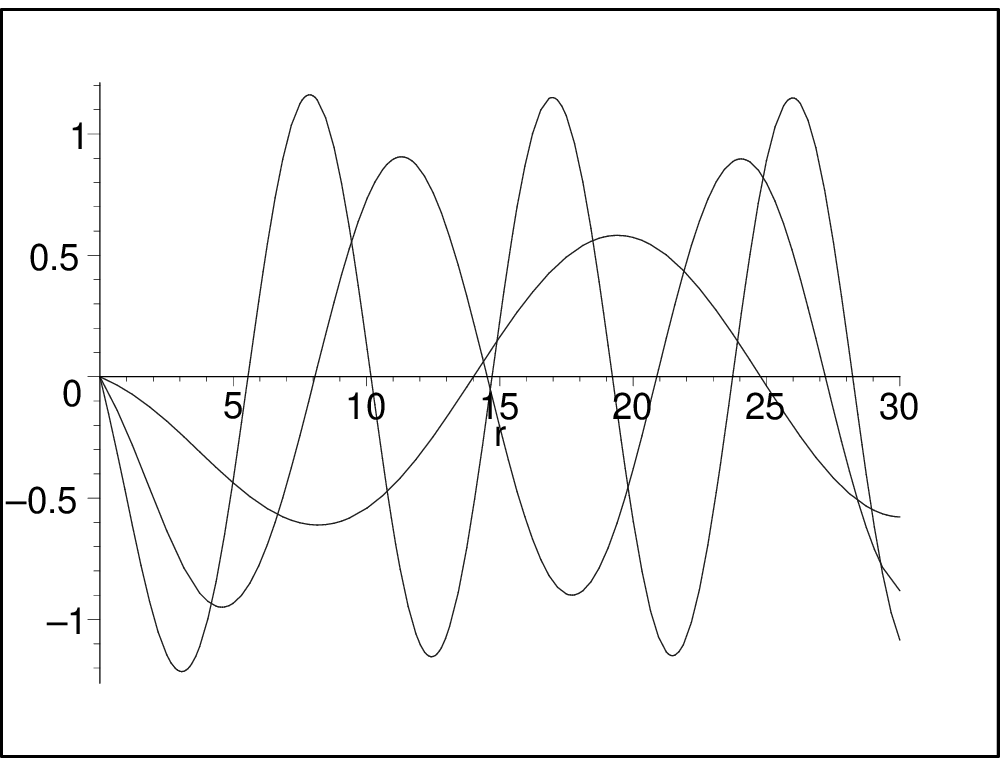}
\includegraphics[width=50mm]{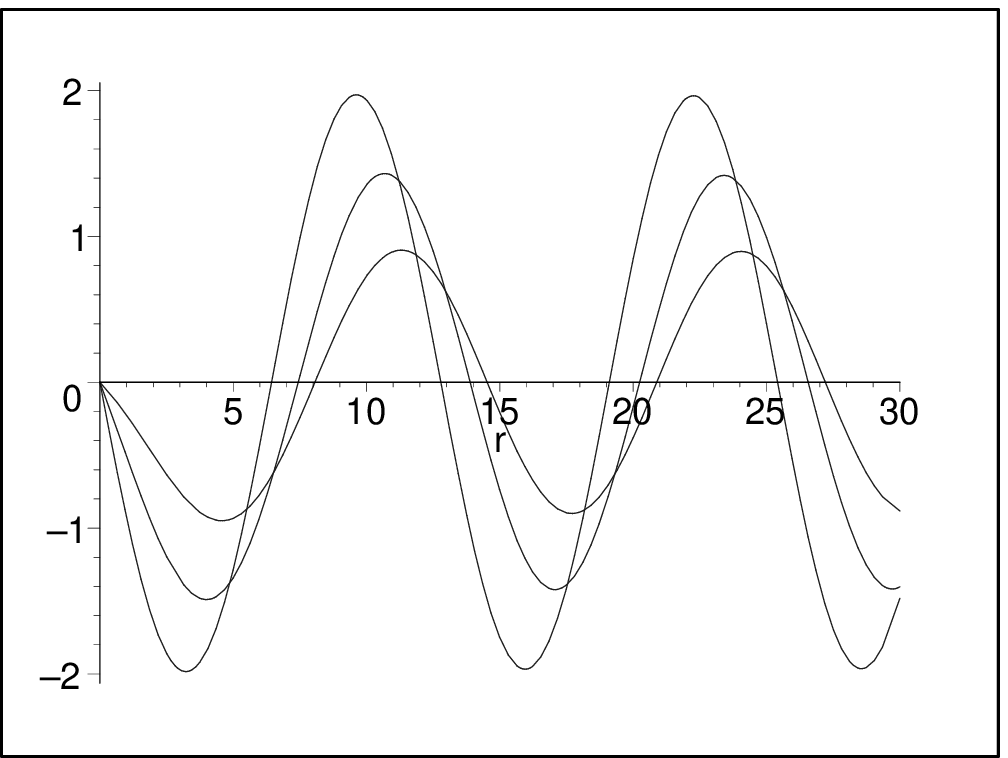}
\includegraphics[width=50mm]{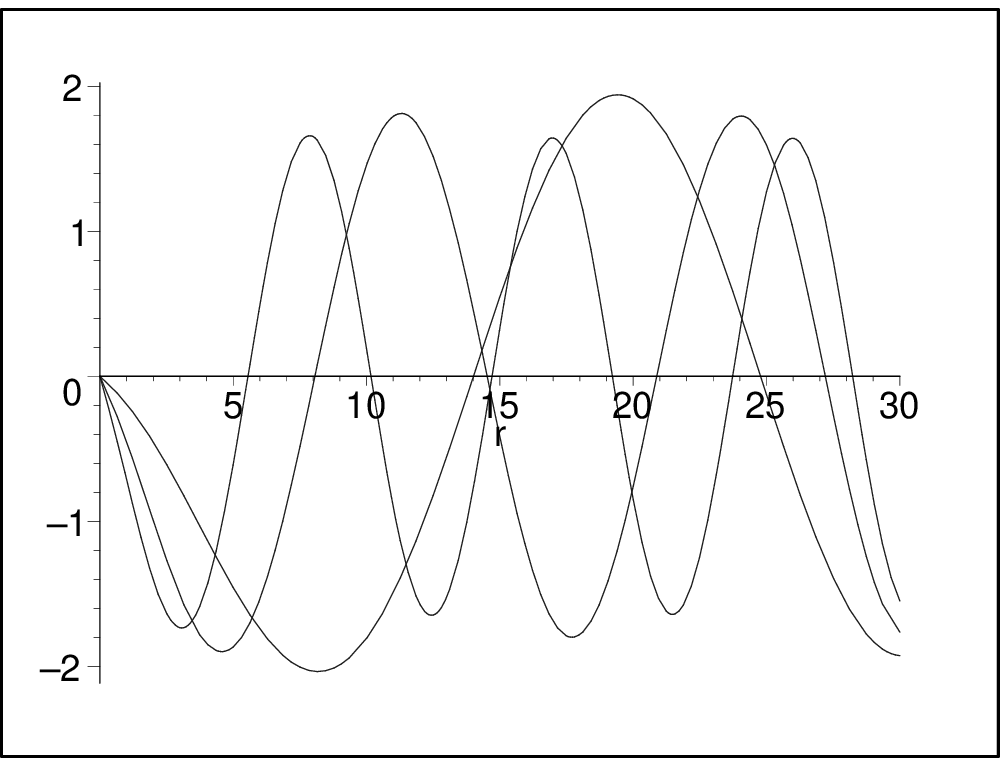}}
\caption{\normalsize
The dependence of the radial wave function (\ref{eq.3.2.1.2}) from
the wave vector $k$ and the parameter $C$
(the values of $\alpha=1$, $r_{0}=0$, $\bar{N}_{2}=1$ are chosen):
(a) a displacement of peaks of the imaginary part of the wave
function along the semiaxis of $r$ is shown with change of the wave
vector $k$
(at $k = 0.3, 0.5, 0.7 $, $C=1 $);
(b) the displacement of the peaks of the imaginary part of the wave
function along the semiaxis of $r$ is shown with change of the
parameter $C$
(at $C = 0.1, 0.5, 1.0$, $k=0.5$);
(c) the real part of the wave function is shown
(at $k = 0.3, 0.5, 0.7$, $C=1$)
\label{fig.3}}
\end{figure}

\begin{figure}[ht]
\centerline{
\includegraphics[width=80mm]{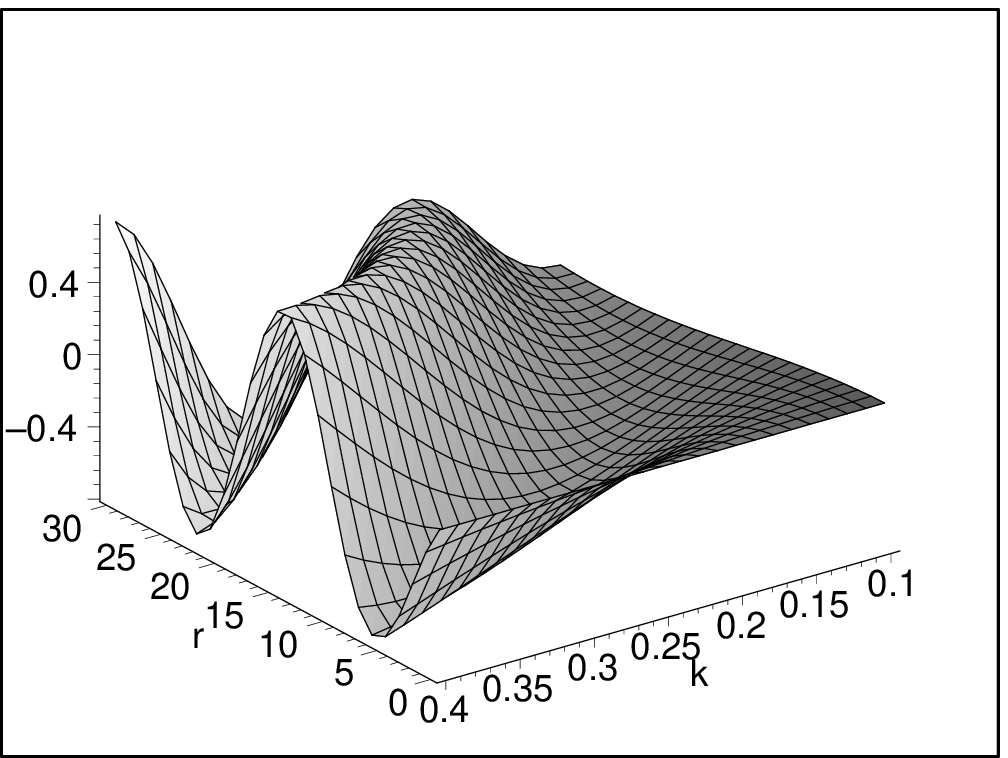}
\includegraphics[width=80mm]{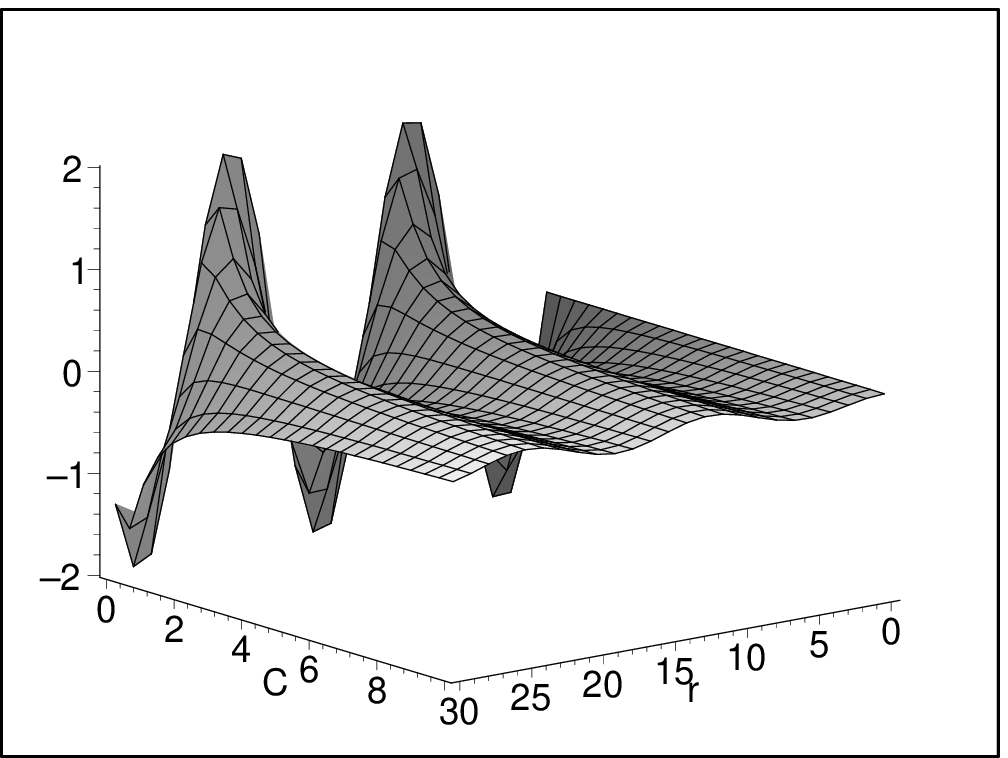}}
\caption{
Behavior of the imaginary part of the wave function
(\ref{eq.3.2.1.2}) depending on the wave vector $k$ and the
parameter $C$
(the values $\alpha=1$, $r_{0}=0$, $\bar{N}_{2}=1$) are used):
(a) dependence on the wave vector $k$ (at $C=1$);
(b) dependence on the parameter $C$ (at $k=0.5$)
\label{fig.4}}
\end{figure}

In Fig.~\ref{fig.3} real and imaginary parts of the wave function
near to the point $r=0$ are shown (here the starting formulas
(\ref{eq.3.2.1.2})--(\ref{eq.3.2.1.3}) are taken). 
From Fig.~\ref{fig.3} (a, b) one can see a deformation of the
imaginary part of this wave function with change of the wave vector
$k$ and the parameter $C$.
The real part of the wave function in its behavior looks like the
imaginary part (see Fig.~\ref{fig.3} (c)). Here, one can see also
that at such choice of the real and imaginary parts of the partial
components of the S-matrix the wave function leaves from its zero
value at $r=0$.
In Fig.~\ref{fig.4} an evident picture of behavior of the imaginary
part of the wave function close to point $r=0$ with continuous change
of the wave vector $k$ and the parameter $C$ is shown.

Note, that according to (\ref{eq.2.4.9}), the condition
(\ref{eq.3.2.1.5}) can bring to not zero values of the radial wave
function $\chi_{l=0}^{(1)}(k,r)$ at $r \to 0$ and can give
discontinuity of the total wave function.
However, a variation of the phase of $S_{l=0}^{(1)}$ does not change
the form of the potential $V_{1}(r)$, which remains zero and
reflectionless.
In other words, the reflectionless potential $V_{1}(r)$ allows an
arbitrariness in a choice of boundary conditions for the wave
function at point $r=0$, and the chosen boundary conditions define
the shape of the total wave function and a process proceeding in the
field of the potential $V_{1}(r)$.
There is a similar situation for the potential $V_{2}(r)$, which
remains reflectionless with the variation of the S-matrix phase.

Now let's analyze the form of the wave function (\ref{eq.3.2.1.2})
in asymptotic region. According to (\ref{eq.3.1.1}), $W(r) \to 0$ at
$r \to +\infty$ and we obtain:
\begin{equation}
  \chi_{l=0}^{(2)}(k,r) =
    \bar{N}_{2} \biggl(e^{-ikr} - S_{l=0}^{(2)} e^{ikr}\biggr).
\label{eq.3.2.1.8}
\end{equation}
One can see, that two components $\chi_{l=0}^{(\pm)}(k,r)$ in
(\ref{eq.3.2.1.2}) represent convergent and divergent waves, that
can be useful for analysis of propagation of the particle in the
field $V_{2}(r)$.
Thus, we have found an \emph{exact analytical division of the total
radial wave function into its convergent and divergent components}
(as for regular and singular Coulomb functions for the known Coulomb
potential) in the description of scattering of the particle in the
inverse power potential (\ref{eq.3.1.5}).

If for the convergent and divergent waves to define radial flows as:
\begin{equation}
  j^{\pm} (k,r) =
    \displaystyle\frac{i\hbar}{2m}
    \biggl( \chi_{l=0}^{(\pm)}(k,r)
      \displaystyle\frac{d \chi_{l=0}^{(\pm), *}(k,r)}{dr}-
      \chi_{l=0}^{(\pm), *}(k,r)
      \displaystyle\frac{d \chi_{l=0}^{(\pm)}(k,r)}{dr} \biggr),
\label{eq.3.2.1.9}
\end{equation}
then for both waves we obtain coincided absolute values of their
flows:
\begin{equation}
  j^{\pm} (k,r) = \pm\displaystyle\frac{\hbar k}{m} |\bar{N}_{2}|^{2}.
\label{eq.3.2.1.10}
\end{equation}
We see, that the flows do not vary in dependence on $r$, and this
gives a fulfillment of a conservation law for the flows from each
wave and the total flow.
Therefore, the convergent wave $\chi_{l=0}^{(-)}(k,r)$ propagates
into the center without the smallest reflection by the field, because
it is defined and is continuous on the whole region of the definition
of the potential (\ref{eq.3.1.5}) and it forms the constant radial
flow $j^{-}(r)$. Now we can tell with confidence, that the
\emph{inverse power radial potential (\ref{eq.3.1.5}), for which we
have found the radial wave function
(\ref{eq.3.2.1.2})--(\ref{eq.3.2.1.4}) for scattering, is
reflectionless at $l=0$}.

Further, one can find the radial wave functions at $l \ne 0$ on the
basis of the same analysis, if for the radial wave function
(\ref{eq.3.2.1.1}) for the potential with zero value to use spherical
Hankel functions instead of factors $\exp{(\pm ikr)}$.

\subsubsection{Wave functions for the reflectionless potential with
the barrier
\label{sec.3.2.2}}

One can use Exp.~(\ref{eq.3.1.4}) for calculation of a new
reflectionless potential $V_{2}(r)$ with a barrier on the basis of
the known reflectionless inverse power potential $V_{1}(r)$. Let's
assume, that these potentials are connected with one superpotential
$W_{2}(r)$.
Let's consider the wave function for the reflectionless inverse power
potential $V_{1}(r)$ at $l=0$ in the form:
\begin{equation}
  \chi_{l=0}^{(1)}(k,r) =
    \bar{N}_{1} \Bigl(f^{-}(r) e^{-ikr} -
    S_{l=0}^{(1)} f^{+}(r) e^{ikr} \Bigr).
\label{eq.3.2.2.1}
\end{equation}
Then the radial wave function at $l=0 $ for the reflectionless
potential $V_{2}(r)$ with the barrier can be found on the basis of
the second expression of (\ref{eq.2.3.6}). Taking into account
(\ref{eq.2.1.4}) and (\ref{eq.2.4.9}), we obtain:
\begin{equation}
\begin{array}{lcl}
  \chi_{l=0}^{(2)}(k,r) & = &
      \displaystyle\frac{\bar{N}_{1}}{N_{2}}
      \biggl( \alpha\displaystyle\frac{d}{dr} + W_{2}(r) \biggr)
      \Bigl(f^{-}(r) e^{-ikr} - S_{l=0}^{(1)} f^{+}(r) e^{ikr} \Bigr) = \\

    & = &
      \displaystyle\frac{\bar{N}_{1}}{N_{2}}
      \biggl[
        \biggl(\alpha \displaystyle\frac{d f^{-}(r)}{dr} -
               ik\alpha f^{-}(r) + W_{2}(r) f^{-}(r) \biggr) e^{-ikr} - \\
    & - &
        S_{l=0}^{(2)}
        \biggl(\alpha \displaystyle\frac{d f^{+}(r)}{dr} +
               ik\alpha f^{+}(r) + W_{2}(r) f^{+}(r) \biggr) e^{ikr}
      \biggr].
\end{array}
\label{eq.3.2.2.2}
\end{equation}
In this expression one can see the division of the total radial wave
function into the convergent and divergent components, that can be
interesting in analysis of scattering (with possible tunneling) of
the particle in the field of the reflectionless potential $V_{2}(r)$
with the barrier.

So, if to use the potential (\ref{eq.3.1.5}) as the first
reflectionless inverse power potential, then we find:
\begin{equation}
  \beta_{2} = 2 \alpha
\label{eq.3.2.2.3}
\end{equation}
and
\begin{equation}
\begin{array}{lcl}
  f^{\pm}(r) =
      1 \pm \displaystyle\frac{i}
            {k \biggl(\bar{r} + \displaystyle\frac{1}{C\alpha} \biggr)}, &
  \displaystyle\frac{d f^{\pm}(r)}{dr} =
      \mp \displaystyle\frac{i}
          {k \biggl(\bar{r} +
          \displaystyle\frac{1}{C\alpha}\biggr)^{2}}, &
  W_{2}(r) =
      \displaystyle\frac{\alpha}{\bar{r}}
      \displaystyle\frac{1 - 6C\alpha\bar{r}^{3}}
                        {1 + 3C\alpha\bar{r}^{3}}.
\end{array}
\label{eq.3.2.2.4}
\end{equation}
Substituting these expressions into (\ref{eq.3.2.2.2}), one can find
the total radial wave function for the reflectionless potential with
the barrier. The value of the partial component of the S-matrix
$S_{l=0}^{(2)}$ can be found from a boundary condition of this wave
function at point $r=0$, as it was made in the previous paragraph for
the inverse power reflectionless potential (\ref{eq.3.1.5}).

\section{Conclusions
\label{sec.conclusions}}

In finishing we note main conclusion and new results.
\begin{itemize}
\item
The new exactly solvable radial reflectionless potential with
barrier, which in the spatial semiaxis of radial coordinate $r$ has
one hole and one barrier, after which it falls down monotonously to
zero with increasing of $r$, is proposed. 
It has shown, that at its shape such potential looks qualitatively
like radial scattering potentials in two-partial description of
collision between particles and nuclei or radial decay potentials in
the two-partial description of decay of compound spherical nuclear
systems.

\item
The found reflectionless potential with the barrier depends on
parameters $\gamma_{n}$ and $C$. One can deform the shape of this
potential: by discrete values of $\gamma_{n}$ (from the sequence
(\ref{eq.3.1.7})) and by continuous values of $C$.
The parameter $\gamma_{n}$ at its variation does not displace visibly
a maximum of the barrier and a minimum of the hole along the semiaxis
$r$, but it changes their absolute values. The parameter $C$ allows
to displace continuously both the barrier maximum and the hole
minimum.

\item
A new approach for construction of a hierarchy of the radial
reflectionless potentials with barriers is proposed.

\item
An exact analytical form for the total radial wave function, its
convergent and divergent components (as for regular and singular
Coulomb functions for the known Coulomb potential) has found in the
description of scattering of a particle in the field of the inverse
power reflectionless potential and in the field of the
reflectionless potential with the barrier (at $\beta=2\alpha$).

\item
It has shown for the inverse power potential, that the radial flows
for the convergent and divergent components of the radial wave
function are constant on the whole semiaxis of $r$, have opposite
directions and coincide by absolute values.
This proves the reflectionless property of the inverse power
potential (with a possible tunneling near the point $r=0$) on the
whole semiaxis $r$. Such analysis is applicable for the found
potential with the barrier also.

\end{itemize}

The analysis has shown, that any selected region of the
reflectionless potential with the barrier (with take into account
both the barrier region, and the small vicinity near $r=0$) does not
influence on the propagation of the particle.
During scattering in the spherically symmetric field with such radial
potential, the particle propagates through it without the smallest
reflection and without any change of angle of direction of its motion
(or tunneling).
One can conclude, that the found radial potential with the barrier is
reflectionless for the propagation of the particle with any kinetic
energy.
If to use it for the two-partial description of the scattering of the
particle on the nucleus with the spherical shape, then one can
conclude, that such nucleus shows itself as \emph{invisible} for the
incident particle.

\bibliographystyle{h-physrev4}
\bibliography{Ref_IMe}


\end{sloppypar}
\end{document}